%% file: nova_qeh_prl_v7.tex
\documentclass[aps,prl,reprint, preprintnumbers, superscriptaddress]{revtex4-1}

\usepackage{graphicx}       
\usepackage{dcolumn}       
\usepackage{bm}               
\usepackage{lineno}          
\usepackage{amsmath}
\usepackage{xspace}
\usepackage{sidecap}
\usepackage{color}           
\usepackage{comment}    
\usepackage{multirow}
\usepackage{hyperref}     
\usepackage{verbatim}
\usepackage{url}
\usepackage{cleveref}
\usepackage{xcolor}
\usepackage{SIunits}

\begin{document}
\preprint{FERMILAB-PUB-26-0385-PPD}
\title{
    Measurement of the $\bar \nu_\mu-$Hydrogen Charged-Current \\ Quasi-Elastic Cross Section using the NOvA Near Detector
}

\input{novaHelastic2026}

\date{\today}

\begin{abstract}
We report a measurement of the total cross section for muon antineutrino charged-current quasi-elastic scattering on hydrogen, 
$\bar \nu_\mu {\rm H} \to \mu^+ n$, in the NOvA near detector using a $1.2\times10^{21}$ proton-on-target exposure in the NuMI beam.  
A selection based on topological and kinematic constraints yields 35,509 signal events in the hydrogen-rich ($10.8\%$) detector, providing the 
highest statistics of (anti)neutrino--hydrogen interactions measured to date. Backgrounds from (anti)neutrino interactions on heavier nuclei are 
constrained using dedicated data control samples, significantly reducing the related systematic uncertainties. We obtain a value 
$\sigma (\bar \nu_\mu {\rm H} \to \mu^+ n) = 0.538 \pm 0.009 ({\rm stat}) \pm 0.010 ({\rm syst}) \pm0.055 ({\rm flux}) \times 10^{-38}$ cm$^2$ 
for the total cross section at an average energy of 1.9 GeV, the most precise total cross-section measurement of this process to date. 
The combined statistical and non-flux systematic uncertainty is more than four times smaller than the flux uncertainty, 
allowing a future use of this measurement to constrain the absolute $\bar \nu_\mu$ flux.
\end{abstract}


\maketitle

\vspace*{-0.40cm} 
{\em Introduction}---%
Current and future long-baseline neutrino experiments~\cite{T2K:2025wet,NOvA:2025tmb,T2K:2023smv,DUNE:2020lwj,Hyper-Kamiokande:2018ofw} 
are pursuing precision measurements of neutrino mixing by comparing neutrino and antineutrino interactions near the source to those after propagation 
over large distances. The detectors at the far site require multi-kton mass with nuclear targets. An understanding of 
(anti)neutrino--nucleus cross sections is therefore essential to extraction of neutrino-oscillation mixing angles, neutrino masses, and leptonic 
CP-violating phase. However, cross sections and flux, smeared by nuclear effects, are folded into the $\nu/\bar\nu$--nucleus interactions 
observed in the detectors. Measurements of $\nu/\bar\nu$ interactions on hydrogen provide a reference free of nuclear effects and 
a probe of the structure of free protons and neutrons. The corresponding nucleon-level amplitudes offer an essential 
input for the modeling of $\nu/\bar\nu$--nucleus cross sections. Furthermore, the same measurements can provide a tool for the determination 
of the (anti)neutrino flux and the energy scale~\cite{Petti:2022bzt,Duyang:2019prb,Petti:2023abz}. 

In this letter, we focus on charged-current quasi-elastic interactions on hydrogen (QEH) $\bar \nu_\mu {\rm H} \to \mu^+ n$, 
characterized by a simple experimental signature. The selection of QEH interactions from CH$_2$ targets using topological and kinematic 
constraints was originally discussed~\cite{Duyang:2018xcc} within the context of the DUNE~\cite{DUNE:2020lwj} near detector. While various 
measurements of antineutrino quasi-elastic-like scattering on nuclear targets were performed 
(for a review see e.g.~\cite{ParticleDataGroup:2024cfk,NOMAD:2009qmu}), extant QEH measurements are scarce. The first observation was made 
using the BNL 7-foot bubble chamber with a liquid-H$_2$ fill, providing the only measurement of the total QEH cross section~\cite{Fanourakis:1980si}, 
based on $13 \pm 6$ events. A more recent detection of QEH interactions on the hydrogen atoms within a plastic scintillator (CH) target was reported by 
MINERvA and used to constrain the nucleon axial form factor and radius~\cite{MINERvA:2023avz}.
The NOvA near detector (ND), with its sizable fiducial mass and relatively large ($10.8\%$) H content from liquid scintillator (mostly CH$_2$) 
and PVC, offers an opportunity for a high-statistics study of QEH interactions. 
Furthermore, NOvA's flux-averaged energy of 1.9 GeV lies between those of the BNL and MINERvA measurements. 
 
{\em Beam and Detector}---%
The NOvA experiment measures neutrino flavor oscillations~\cite{NOvA:2025tmb} using two detectors separated 
by \unit{810}{\kilo \meter}, placed \unit{14.6}{\milli \radian} off-axis from the central direction of the Fermilab NuMI beam~\cite{ref:Adamson:2015dkw}. 
Magnetic focusing horns in the beamline charge-select neutrino parents yielding a flux comprising 93\% $\bar \nu_\mu$, 6\% $\nu_\mu$, 
and less than 1\% $\nu_e + \bar \nu_e$ in the antineutrino beam mode between 1 and \unit{5}{\giga \electronvolt}. 
The average energy is about \unit{1.9}{ \giga \electronvolt } at the ND site. This letter reports results from data collected in the ND 
from June 2016 to July 2019, equivalent to $1.2 \times 10^{21}$ protons on target (POT). 

The ND is a tracking calorimeter composed of liquid scintillator contained in rows of cells within planar PVC extrusions. 
The fiducial volume of the detector relevant for this measurement is \unit{3.3}{\meter}~$\times$~\unit{3.1}{\meter}~$\times$~\unit{11.5}{\meter}. 
The detector is segmented according to the PVC cell cross section, measuring \unit{6.6}{\centi \meter} along the beam direction (15\% of a radiation length)
and \unit{3.9}{\centi \meter} transverse to it. The cell length spans either the height or width of the detector in planes alternating between horizontal 
and vertical position measurements. Each cell is filled with a blend of 95\% mineral oil (CH$_2$) and 5\% pseudocumene with 
trace concentrations of wavelength-shifting fluors~\cite{ref:scintillator}. The resulting composition by mass is 66.7\% carbon, 16.1\% chlorine, 
10.8\% hydrogen, 3.2\% titanium, 3.0\% oxygen with other trace elements. The total target mass within the fiducial volume is 116.7 tons, 
of which 12.5 tons is hydrogen. 

A charged particle passing through the detector cells deposits energy and produces scintillation and Cherenkov light, which is captured by 
wavelength-shifting fibers and transferred to an avalanche photodiode (APD) at one end of each cell. The APD output is digitized using 
custom-made front-end electronics. All signals above a noise-vetoing threshold ($\sim$0.3 MIP) are sent to a data buffer. A downstream muon 
detector is constructed from pairs of planes separated by slabs of steel and is designed to measure muons with energies up to 2.5 GeV by range. 

{\em Simulations}---%
The NuMI flux is predicted using Geant4 v4.10~\cite{AGOSTINELLI2003250} with the {\rm FTFP\_BERT} hadronic model, which is 
reweighted using the Package to Predict the Flux (PPFX)~\cite{Aliaga:2016oaz} to incorporate constraints from hadron production measurements~\cite{Paley:2014rpb,Alt:2006fr,Abgrall:2011ae,Barton:1982dg,Seun:2007zz,Tinti:2010zz,Lebedev:2007zz,Baatar:2012fua,Skubic:1978fi,Denisov:1973zv,Carroll:1978hc,Abe:2012av,Gaisser:1975et,Cronin:1957zz,Allaby:1969de,Longo:1962zz,Bobchenko:1979hp,Fedorov:1977an,Abrams:1969jm}. 
Charged (CC) and Neutral Current (NC) $\nu/\bar \nu$ interactions in the ND materials are simulated with the GENIE v3.0.6~\cite{GENIE:2021zuu} 
package in a customized {\rm G18\_10j\_00\_000} configuration. 
The QEH process is simulated with the Llewellyn-Smith model~\cite{LlewellynSmith:1971uhs}, replacing the dipole axial form factor with the 
more refined description afforded by the $z$-expansion model~\cite{Meyer:2016oeg}. The initial states of the nucleons within the various target 
nuclei are generated with the local Fermi Gas model~\cite{Nieves:2011pp}. Quasi-elastic (QE) and meson-exchange current (MEC) interactions 
in nuclei are simulated using the Valencia local Fermi gas model~\cite{Gran:2013kda}, resonant (RES) and coherent (COH) meson production 
with the Berger--Sehgal model~\cite{Berger:2007rq}, and deep-inelastic scattering (DIS) with the Bodek--Yang model~\cite{Bodek:2002vp}. 
The final-state interactions (FSI) of primary hadrons in the nucleus are simulated with the GENIE hN2018 model~\cite{Andreopoulos:2015wxa}. 

The GENIE output is reweighted by adjusting the MEC and FSI models to obtain central values and uncertainties consistent with the inclusive 
CC interactions observed in the NOvA ND~\cite{NOvA:2020rbg}. 
The MEC weights are parametrized as a function of the energy and momentum transfer of the event~\cite{NOvA:2024zmr}. 
The FSI weights and related uncertainties are determined using Boosted Decision Trees (BDT) to better agree with pion 
scattering data~\cite{Allardyce:1973ce,Saunders:1996ic,Meirav:1988pn,Levenson:1983xu,Ashery:1981tq,Ashery:1984ne,DUET:2016yrf}. 
Only inclusive CC data collected with the neutrino beam mode are used to determine the weights, which are then 
applied without modification to the data collected in the antineutrino beam mode relevant for the present analysis. 

Geant4 v4.10~\cite{AGOSTINELLI2003250} with the {\rm QGSP\_BERT\_HP}  configuration is used to simulate the interaction of and energy 
deposition by final-state particles within the detector. The production and transfer of scintillation and Cherenkov light in the detector cells to the 
readout electronics are simulated using custom software~\cite{Aurisano:2015oxj}. 

\begin{figure}[!b]
\centering
\includegraphics[width=\columnwidth]{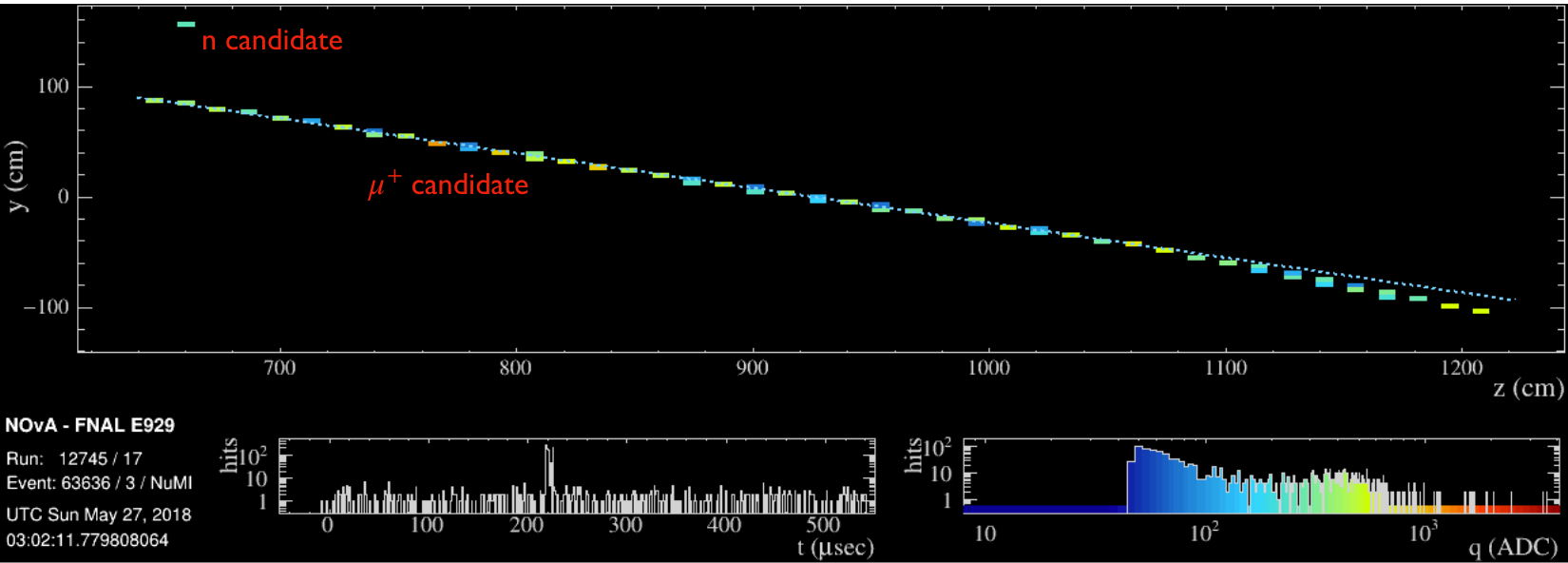}
\caption{Example of selected QEH candidate from NOvA data. The contained event lies within the signal region and is characterized by high values of both 
HitID and KineID (see text).}
\label{fig:NOvA_QEH-CandidateQEH}
\end{figure}

{\em Event Selection}---%
The signature of QEH interactions is a single muon track reconstructed in the detector, associated with an isolated cluster of hits from the 
daughter particles created by the neutron interacting in the detector materials (Fig.~\ref{fig:NOvA_QEH-CandidateQEH}). The backgrounds are from 
all types of (anti)neutrino interactions with the heavier nuclei within the composite target. Particle trajectories (tracks) are obtained by grouping hits 
correlated in space and time via a Kalman filter-based algorithm in both the horizontal and vertical two-dimensional detector views~\cite{Baird:2015pgm}. 
We require the presence of only one three-dimensional Kalman track in the event, formed by combining tracks 
from the two views based on their overlap in the longitudinal direction. No explicit cut nor muon identification criteria are applied to the selected track, 
since the kinematic selections described later guarantee a muon purity of 99.9\%. The average muon momentum resolution from range is 
about 3.5\% for tracks stopping within the detector. The resolution on the muon angle with respect to the beam $\theta_\mu$ is about 14 mrad. 

\begin{figure}[!t]
\centering
\includegraphics[width=\columnwidth]{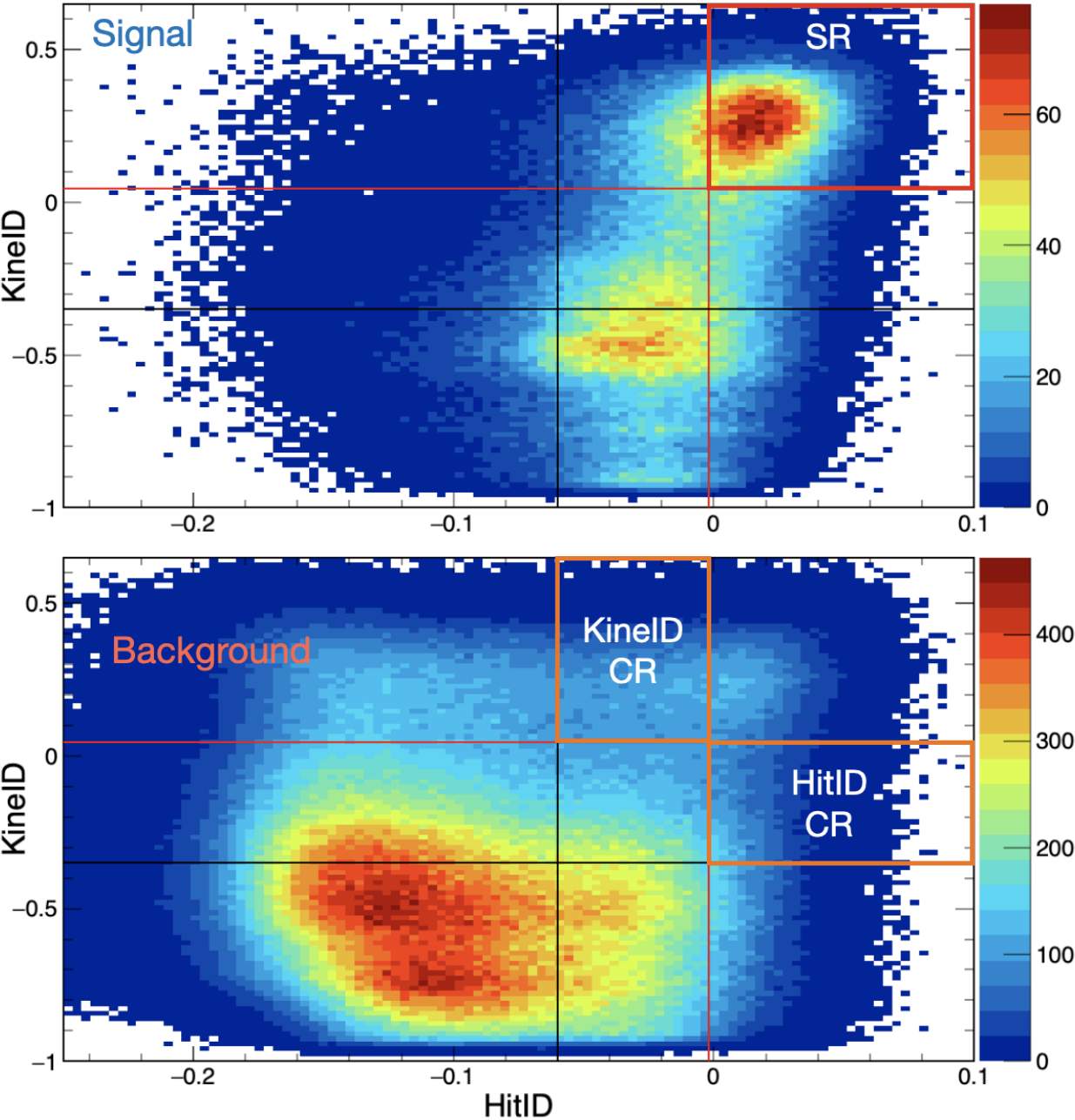}
\caption{Definition of the signal region (SR) and control regions (CR) in the two-dimensional plane of the topological (HitID) and kinematic (KineID) 
multivariate discriminants. The distributions shown are for contained events and are normalized to $1.2 \times 10^{21}$ POT.}
\label{fig:QEH-2DplaneNN}
\end{figure}

The reconstructed muon momentum vector is used to calculate the expected neutron momentum vector based on energy-momentum 
conservation for a proton target at rest (natural units): 
\begin{eqnarray}
E_n & = & \frac{M_n^2-m_\mu^2+2M_pE_\mu -M_p^2}{2\left(  M_p- E_\mu + {p}_\mu  \cos \theta_\mu\right)}  +M_p-E_\mu \, , \\ 
{\bf p}_{\rm T}^n & = & - {\bf p}_{\rm T}^\mu  \, , \\ 
p_{\rm L}^n & = & \frac{M_{\mathrm{n}}^2+(p_{\mathrm{T}}^{\mu})^2-\left(p_{\mathrm{L}}^\mu-E_\mu+M_{\mathrm{p}}\right)^2}{2\left(p_{\mathrm{L}}^\mu-E_\mu+M_{\mathrm{p}}\right)^2} \, , 
\end{eqnarray} 
where $M_p, M_n, m_\mu$ are the masses of the proton, neutron, and muon, respectively, ${\bf p}_{\mu (n)}$ and $E_{\mu (n)}$ 
are the momentum vector and energy of the outgoing muon (neutron), and the suffix ${\rm T} ({\rm L})$ refers to 
the corresponding transverse (longitudinal) components. 

Clusters of fewer than 6 hits disconnected from the Kalman track are considered as potential neutron candidates. 
For each event a single neutron candidate is selected as the closest one to the calculated line-of-flight of the neutron. 
Most of the resulting neutron candidates ($\sim 97\%$) consist of a single hit. No explicit cut is applied on them. The average neutron 
purity is 82.3\% with an efficiency of 55.5\%, shown in the supplemental material as a function of the kinetic energy. The subsequent kinematic 
selections further increase the neutron purity up to about 98.1\% in the final samples. 

The reconstructed neutron momentum vector is obtained from the line connecting the first hit of the muon Kalman 
track (vertex) with the position of the neutron candidate hits, multiplied by the magnitude of the calculated neutron momentum~\cite{Duyang:2018xcc}. 
The only neutron hit information used in the analysis is therefore the location of the neutron interaction point in the detector. 
The average neutron momentum resolution is about 8.4\%, as shown in the supplemental material. 

The events are subsequently partitioned into two complementary samples, which are subject to separate event selections. In the ``contained'' sample  
the muon track has stopped several cells before reaching any detector edge to ensure containment of all the neutrino energy~\cite{NOvA:2021eqi}. 
The ``uncontained'' sample includes all remaining events, in which the muon track may have exited the detector. 

\begin{table*}[!t]
\centering
\caption{Observed and expected events in the SR for the contained and uncontained samples broken down by process. The ratios include 
statistical and systematic (Table~\ref{tab:NOvA_QEH-Systematics}) uncertainties, with the exception of the common flux systematics.}
\begin{tabular}{r|c|c|c|c|c|c|c|c|c|c|c|r} 
Sample & Signal & CCQE &  CCMEC & CCRES & CCDIS & CCCOH & NC & $\nu_\mu,\nu_e,\bar \nu_e$ &  Total Bkg  &   Total &   Data  &     Data/Expected~~~   \\ 
\hline
Contained  & 15,103  & 21,559 & 4,238 & 773  & 221 &  15 & 283 & 751 & 27,839 & 42,941 & 43,644  & 1.016$\pm$0.010$\pm$0.013~ \\ 
Uncontained  & 20,670  & 33,024 & 7,389 & 1,987 & 317 & 19 & 62 & 747 & 43,544 & 64,214 & 63,248 & 0.985$\pm$0.008$\pm$0.013 
\label{tab:NOvA_QEH-BoxOpening}
\end{tabular}
\end{table*}

The selection of QEH events is defined by two cuts on separate multivariate discriminants. The first (HitID) incorporates topological 
information based on the location and energy of the hits/trajectories via 14 input variables (of comparable weights): 
the energy depositions in the first 4 hits of the muon Kalman track; the energy deposited by particles other than the muon in boxes of 10, 15, and 
20 cm half-width centered around the vertex; the energy deposited by particles other than the muon within a sphere of radius equal to the distance 
between the vertex and the neutron interaction point; the ratio between the numbers of muon hits and total hits; the number of isolated clusters; 
the total number of two-dimensional trajectories with a single detector view; the number of trajectories connected to the vertex; the distance of 
closest approach between the neutron candidate and the calculated line-of-flight; and the calculated neutron energy. 

The second discriminant (KineID) describes the global event kinematics used to separate interactions on H 
(at rest without nuclear smearing) from interactions on a nucleon within a nucleus, based on 7 input variables~\cite{Duyang:2018xcc} 
(of comparable weights): the transverse momenta of the neutron and muon,  $p_{\rm T}^n$ and $p_{\rm T}^\mu$; the missing transverse 
momentum $p_{\rm T}^m$; the opening angle between the muon and the neutron $\theta_{\mu n}$; the angle of the total visible momentum with the 
beam direction $\theta_{\nu T}$; the difference between the calculated and measured neutrino energies 
$\Delta E_\nu = E_n+E_\mu-M_p - \mid {\bf p}_\mu + {\bf p}_n \mid$; and the neutron longitudinal momentum $p_{\rm L}^n$. Both discriminants 
are built from BDT-based classifiers, which are trained separately for the contained and uncontained samples using the TMVA package 
in ROOT~\cite{Hocker:2007ht}. 

A signal region (SR) is defined in the plane of the two discriminants (Fig.~\ref{fig:QEH-2DplaneNN}), together with an appropriate control 
region (CR) for each of the two discriminants. The CRs depend on the cuts used for the SR and are an essential feature of the analysis. 
The background statistics within the CRs is required to be larger than the statistics of the SR and the corresponding background compositions 
to be similar. We maximized the purity of the selected QEH samples simultaneously with the optimization of the CRs. 
A QEH purity of about 41\% (36\%) with a signal efficiency $\varepsilon_{\rm QEH}$ of 3.6\% (4.8\%) are obtained for the contained (uncontained) 
sample from the simulations. The corresponding efficiency for the kinematic selection is 58\% (34\%). 
The SR was blinded until the analysis was finalized in order to avoid biases. 

{\em Data Corrections}---%
The CRs are used to obtain data-driven background predictions compensating the limitations of the Monte Carlo (MC) simulations. 
A first correction ($k_{\rm nc}$) is applied to take into account differences in the neutron interaction rate and detection efficiency between data and MC. 
The $k_{\rm nc}$ correction is determined from the ratios of data to MC events at the neutron candidate level, after removing the SR. 
The QEH content is only about 7\% at this stage and is subtracted from the samples. A $k_{\rm nc}$ value of 0.90 (0.93) is obtained for the 
contained (uncontained) sample, and is applied to both backgrounds and signal, since the energy distributions for the selected neutron 
candidate are found to be similar in both (shown in the supplemental material). 
\begin{table}[!b]
\centering
\caption{Fractional systematic uncertainties on the total QEH cross section by various sources in the contained and uncontained samples, 
and respective correlations. 
}
\begin{tabular}{r|c|c|c}
Source & Cont.\ (\%) & Uncont.\ (\%) &  Correlation \\ 
\hline
Normalization & ~0.9  & ~0.9  & ~1.00 \\                     
Background estimate & ~1.2 & ~1.3  &  ~0.96 \\          
QEH modeling & ~0.8  & ~1.3    &  ~0.30  \\                   
Neutron detection & ~1.1  & ~1.1 &  -0.47 \\                 
Muon energy scale & ~0.6  & ~0.2 &  -0.07  \\              
Muon angle scale  & ~0.6 & ~0.2   &  -0.23  \\              
Detector response &   ~1.0  &   ~0.5  &  -0.58   \\         
\hline
Systematics (w/o flux) & ~2.4 & ~2.4  &     \\ 
\hline 
Integrated flux      &  10.2  &   10.2   &  ~1.00   
\label{tab:NOvA_QEH-Systematics}
\end{tabular}
\end{table}

The effect of the sequential cuts on HitID and KineID is then corrected for background events using the 
two CRs separately, after removing their small ($\lesssim 10\%$) signal content. For each cut, we evaluate the efficiency in data and MC from the corresponding 
CR, and use their ratio as correction factor $k_{\rm cut}$ for all background efficiencies~\cite{NOMAD:2001xxt}: 
\begin{equation} 
\varepsilon_{\rm Bkg} = \varepsilon_{\rm MC} \times \varepsilon^{\rm CR}_{\rm Data} / \varepsilon^{\rm CR}_{\rm MC} = \varepsilon_{\rm MC} \times k_{\rm cut} \, , 
\label{eq:DScorr} 
\end{equation} 
where $\varepsilon_{\rm MC}$ is the initial MC efficiency. 
The resulting background predictions are largely insensitive to the details of the MC, up to the statistics of the CRs. 
The value of the HitID correction ($k_{\rm HitID}$) is 1.30 (1.11) for the contained (uncontained) sample, while the value of the KineID 
correction ($k_{\rm KineID}$) is 0.87 (1.01). The combined effect of all three corrections $k_{\rm nc}$, $k_{\rm HitID}$, and $k_{\rm KineID}$ is 1.02 (1.04). 
The statistics of the CRs contributes to the statistical uncertainties of the subtracted backgrounds. 
Table~\ref{tab:NOvA_QEH-BoxOpening} summarizes the corrected backgrounds in the SR broken down by process. 

\begin{figure}[!b]
\centering
\includegraphics[width=\columnwidth]{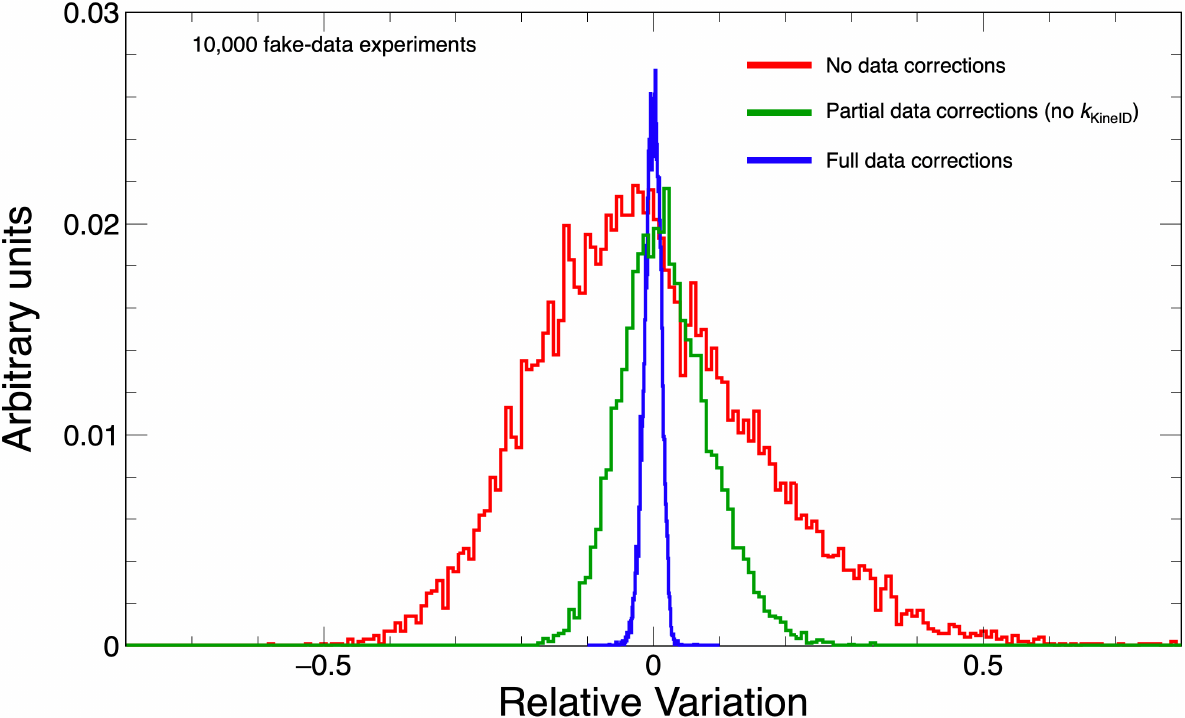}
\caption{Variations in the background predictions for contained events obtained in 10,000 fake-data experiments with random variations in the 
neutrino interaction models, in the neutron propagation model, and in the detector model without data corrections (red), and with partial (green) 
and complete (blue) data corrections. The three histograms are arbitrarily normalized for the sake of comparing their widths.} 
\label{fig:NOvA_QEH-DataSimulator}
\end{figure}

{\em Systematic Uncertainties}---%
Uncertainties on the integrated antineutrino flux (10.2\%) arise from hadron production uncertainties, 
evaluated by the PPFX package~\cite{Aliaga:2016oaz}, and 
from beam optics modeling. The normalization uncertainty is determined by the knowledge of the hydrogen content in the target materials, 
based on data from the manufacturing and construction processes, and of the total POT collected, based on measurements of the beam current 
through a toroid~\cite{NOvA:2021eqi}. Systematic uncertainties on (anti)neutrino interactions with the detector materials are 
estimated for QEH and backgrounds by modifying various parameters in the GENIE event generator and the MC simulations according 
to physics considerations and external data~\cite{NOvA:2021eqi, NOvA:2025tmb}. The neutron detection systematics are determined using an 
alternative neutron propagation model, MENATE~\cite{menate:eurisol,Kohley:2012awa}. 
Uncertainties on detector response are evaluated by varying the parameters of the detector calibration, the production and transport of scintillation 
and Cherenkov light, and the modeling of muon energy loss in the detector materials. The uncertainties on the muon energy and angle 
were evaluated by varying them by 1.2\% and 2.5 mrad, respectively~\cite{NOvA:2021eqi,Strait:2019dya}. 

For each systematic variation, the corresponding uncertainty is assessed by repeating the complete analysis --  including an update of   
the $k_{\rm nc}, k_{\rm HitID}$, and $k_{\rm KineID}$ data corrections -- with the shifted background and signal MC and comparing the results with those 
using the nominal parameters. Table~\ref{tab:NOvA_QEH-Systematics} summarizes the various sources of uncertainty on the total QEH 
cross-section measurements for the contained and uncontained samples, as well as their correlation. All systematic uncertainties pertaining 
to detection and reconstruction effects are anticorrelated in the two samples. 

To evaluate the robustness of the background predictions we simulated 10,000 fake-data experiments in which we randomly varied
the neutrino and neutron interaction models and detector model. For each experiment, we generated 
a random sequence of variations thrown from Gaussian distributions with standard deviations equal to the uncertainties on each parameter. 
Results are shown in Fig.~\ref{fig:NOvA_QEH-DataSimulator}. The data corrections maintain a stable background prediction despite systematic 
variations up to about 60\% in the MC, due to compensation from the corresponding $k_{\rm cut}$ factors in Eq.~\eqref{eq:DScorr}. 
The latter are anticorrelated with the MC variations, as illustrated in the supplemental material. 
The resulting uncertainties are reduced by about an order of magnitude and are consistent with the values listed in Table~\ref{tab:NOvA_QEH-Systematics}. 

Additional checks were performed by dropping all MC weights, as well as by removing the MEC events from the MC samples. In all cases 
we obtained results consistent with the fake-data cross section within the quoted systematic uncertainties.

{\em Results}---%
Table~\ref{tab:NOvA_QEH-BoxOpening} shows the numbers of data events in the SR for the contained and uncontained samples. In both cases we obtain 
good agreement with the corresponding predictions including all data corrections. A total of 35,509 $\pm$ 679({\rm stat}) QEH events are observed after 
background subtraction, against 35,773 expected. This sample represents the highest statistics of (anti)neutrino-hydrogen interactions measured to date. 

The flux-averaged total QEH cross section is determined from the background-subtracted signal events as
\begin{equation}
\sigma_{\rm QEH} = \frac{N_{\rm Data} - N_{\rm Bkg}}{\varepsilon_{\rm QEH} \; \Phi \; N_{\rm H}} \, , 
\end{equation} 
\noindent where $N_{\rm Data}$ is the content of the signal region, $N_{\rm Bkg}$ the data-based background estimate, $\Phi$ the integrated flux, and $N_{\rm H}$ the number of H atoms in the fiducial volume. The resulting values of $\sigma_{\rm QEH}$ 
are $0.565\pm0.014({\rm stat})\pm0.013({\rm syst})\pm0.058({\rm flux}) \times 10^{-38}$ cm$^2$ and 
$0.515\pm0.013({\rm stat})\pm0.013({\rm syst})\pm0.052({\rm flux}) \times 10^{-38}$ cm$^2$ for the contained and uncontained samples, respectively. 

\begin{figure}[!t]
\centering
\includegraphics[width=\columnwidth]{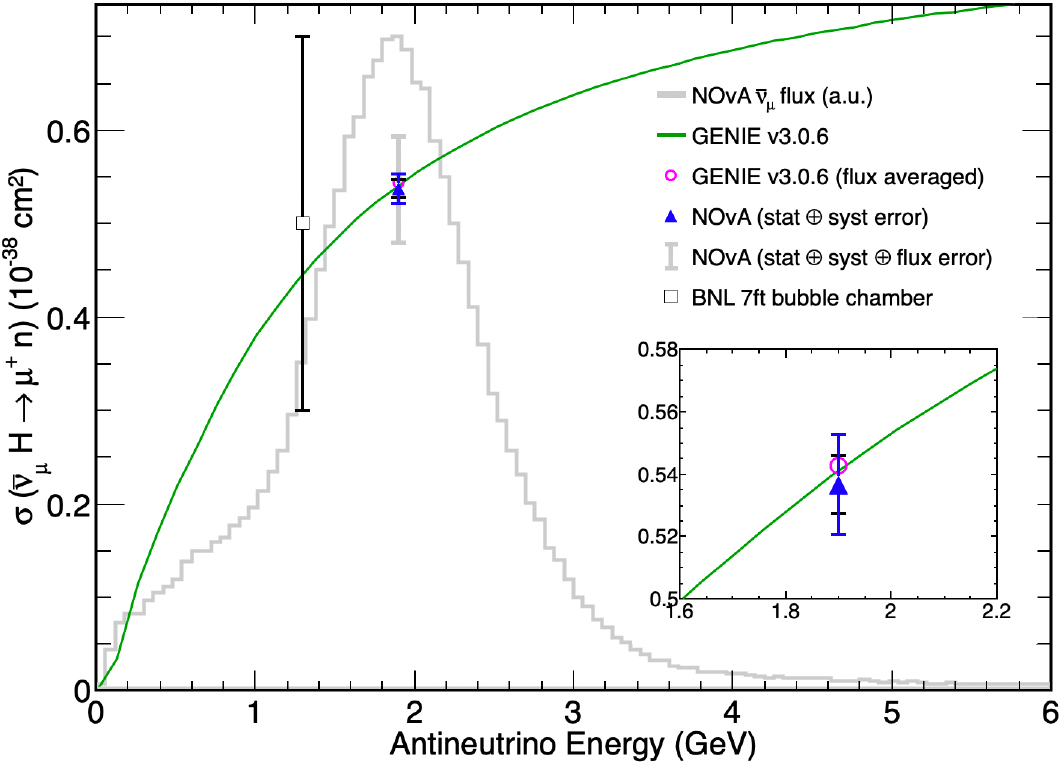}
\caption{Flux-averaged total cross section for the exclusive process $\bar \nu_\mu {\rm H} \to \mu^+ n$ obtained from the combination of the contained and 
uncontained samples. The combined statistical and systematic uncertainty (shown separately in the inset) is compared with the uncertainty on the 
integrated flux. The only existing measurement~\cite{Fanourakis:1980si} of the total cross section is also shown.}
\label{fig:NOvA_QEH-ResultsXS}
\end{figure}

The contained and uncontained samples are mutually exclusive and provide complementary information covering a broad phase space. 
While the selection efficiency is larger  at lower (higher) neutrino and muon energies in the contained (uncontained) sample, the combined efficiency 
and purity are rather uniform across kinematic variables (shown in supplemental material). The two samples are also characterized by a somewhat 
different mixture of systematic uncertainties with varying correlations  (Table~\ref{tab:NOvA_QEH-Systematics}), allowing a more accurate 
control of systematics. We therefore combine them into our final result for the total QEH cross section measured at the flux-averaged energy of 1.9 GeV: 
\begin{equation} 
0.538 \pm 0.009 ({\rm stat}) \pm 0.010 ({\rm syst}) \pm 0.055 ({\rm flux}) \times 10^{-38} {\rm cm}^2 \, ,
\label{eq:xsec} 
\end{equation} 
where the non-flux systematic uncertainties (1.9\%) are obtained from the best linear unbiased estimator~\cite{huang1930probability} taking into account 
the correlation between the contained and uncontained samples for each individual source listed in Table~\ref{tab:NOvA_QEH-Systematics}. 
This result is the most precise measurement of the total QEH cross section to date. Figure~\ref{fig:NOvA_QEH-ResultsXS} compares  
the NOvA measurement in Eq.~\eqref{eq:xsec} with the only existing measurement~\cite{Fanourakis:1980si}, from the BNL 7-foot bubble chamber, 
together with the spectrum of the NOvA $\bar \nu_\mu$ flux. The expectation from the GENIE v3.0.6~\cite{GENIE:2021zuu} event generator, 
based on the Llewellyn-Smith model~\cite{LlewellynSmith:1971uhs} with the dipole axial form factor replaced by the $z$-expansion 
model~\cite{Meyer:2016oeg}, is also shown. The combined statistical and non-flux systematic uncertainty (2.5\%) is more than four times smaller 
than the uncertainty on the integrated flux, allowing a future use of this measurement to constrain the absolute 
$\bar \nu_\mu$ flux~\cite{Petti:2023abz,Duyang:2019prb}.  \nocite{Talukdar:2026} 
The data related to this measurement, their uncertainties, and the antineutrino flux can be found at \cite{ref:data_release}.

{\em Acknowledgements}---%
This document was prepared by the NOvA collaboration using the resources of the Fermi National Accelerator Laboratory (Fermilab), a U.S. Department of Energy, Office of Science, HEP User Facility. Fermilab is managed by Fermi Forward Discovery Group, LLC, acting under Contract No. 89243024CSC000002.  This work was supported by the U.S. Department of Energy; the U.S. National Science Foundation; the Department of Science and Technology, India; the European Research Council; the MSMT CR, GA UK, Czech Republic; the RAS, the Ministry of Science and Higher Education, and RFBR, Russia; CNPq and FAPEG, Brazil; UKRI, STFC and the Royal Society, United Kingdom; and the state and University of Minnesota.  We are grateful for the contributions of the staffs of the University of Minnesota at the Ash River Laboratory, and of Fermilab. For the purpose of open access, the author has applied a Creative Commons Attribution (CC BY) license to any Author Accepted Manuscript version arising.

\bibliographystyle{apsrev4-1}
\bibliography{nova_qeh_prl.bib}

\clearpage
 
\onecolumngrid
\section{Supplemental Material for the Measurement of the $\bar \nu_\mu-$Hydrogen Charged-Current Quasi-Elastic Cross Section using the NOvA Near Detector}

\setcounter{figure}{0}
\setcounter{table}{0}
\setcounter{page}{1}
\setcounter{section}{0}
\setcounter{secnumdepth}{4}
\makeatletter
\renewcommand{\theequation}{S\arabic{equation}}
\renewcommand{\thefigure}{S\arabic{figure}}
\renewcommand{\thetable}{S\arabic{table}}
\renewcommand{\thepage}{S\arabic{page}}
\renewcommand{\thesection}{S-\Roman{section}}

\begin{figure}[!h]
\centering
\includegraphics[width=1.00\linewidth]{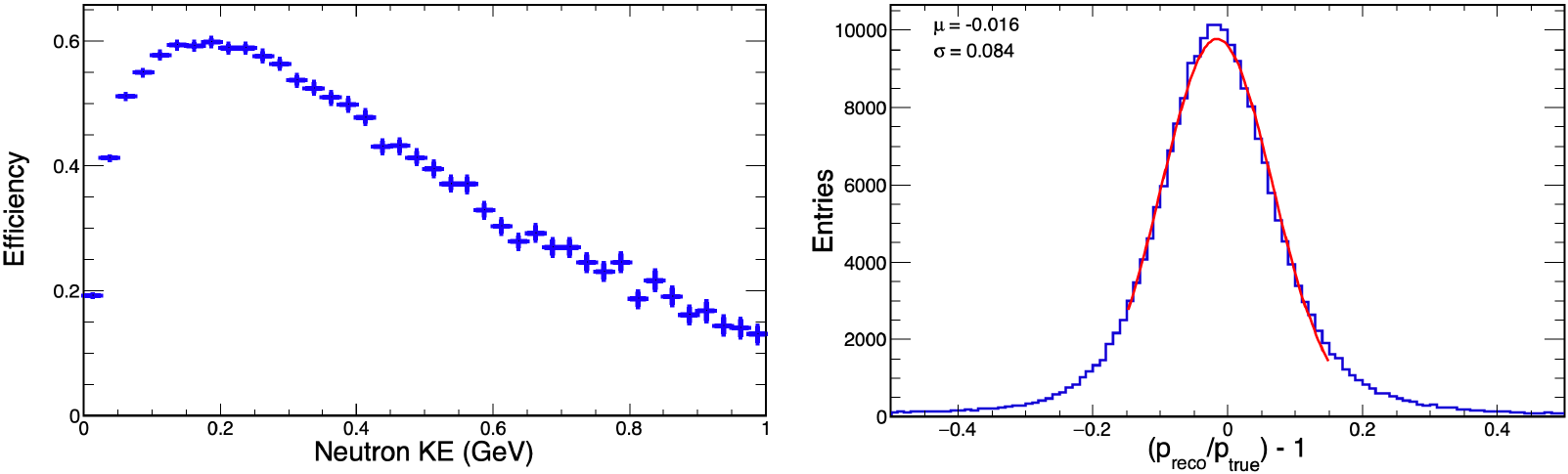}
\caption{(Left) Neutron detection efficiency as a function of the kinetic energy (KE) for QEH signal events in the combined contained and uncontained samples. 
(Right) Resolution on the reconstructed neutron momentum for QEH signal events with the muon stopping in the detector.}
\label{fig:NOvA_QEH-NeutronRec}
\end{figure}

\begin{figure}[!h]
\centering
\includegraphics[width=0.50\linewidth]{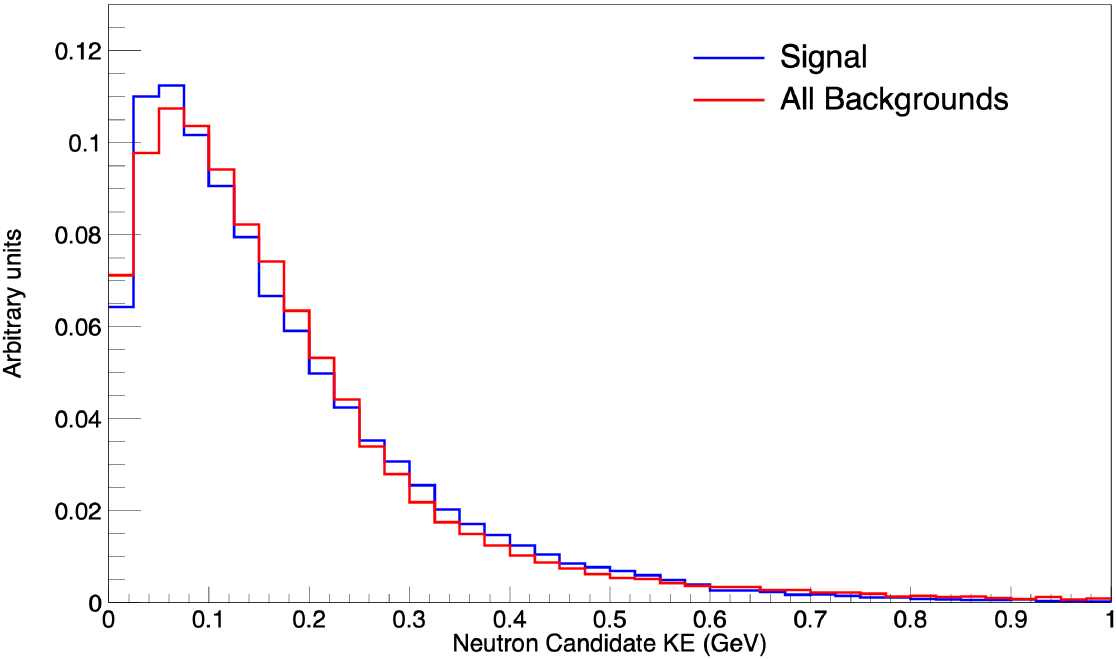}
\caption{Distributions of the kinetic energy for the selected neutron candidate in QEH signal and background events for the combined contained and uncontained samples. The distributions are normalized to unit area.}
\label{fig:NOvA_QEH-NeutCandKE}
\end{figure}

This supplemental material provides additional details supporting the main results presented in the Letter. 
The neutron reconstruction and the selection of the QEH neutron candidate are described in the main body of the paper.  
Figure~\ref{fig:NOvA_QEH-NeutronRec} shows the neutron detection efficiency as a function of the kinetic energy, as well as the resolution on the neutron 
momentum for QEH signal events with the muon stopping in the detector. For interactions with nuclei more than one neutron can be produced, 
resulting on average in a lower kinetic energy compared to the single neutron produced in QEH events. However, our selection procedure of a single 
neutron for each event biases its spectrum towards signal-like configurations. The distributions of the kinetic energy for the selected neutron candidate 
are thus similar in both signal and background events, as illustrated in Fig.~\ref{fig:NOvA_QEH-NeutCandKE}. 


\begin{figure}[!t]
\centering
\includegraphics[width=0.50\linewidth]{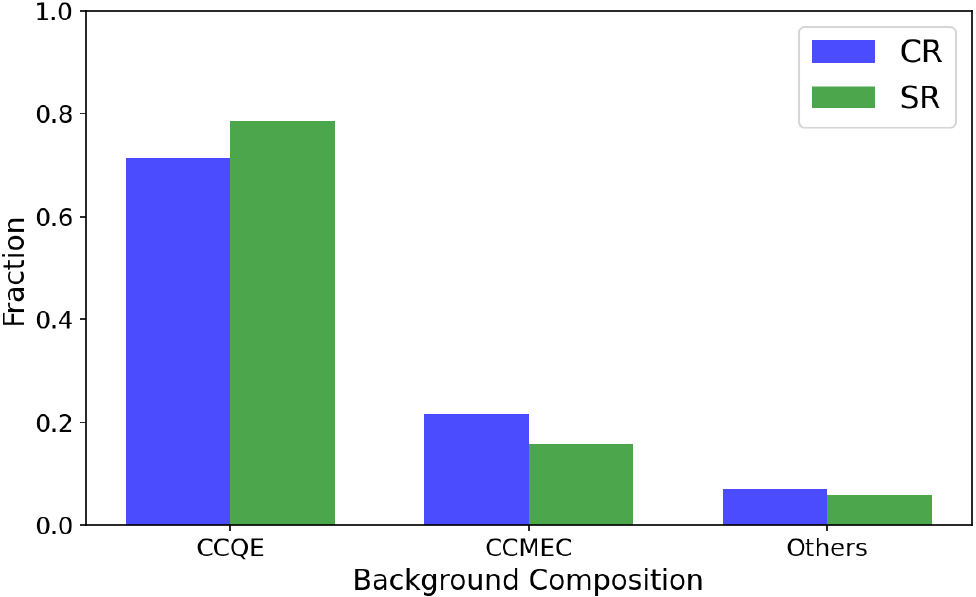}
\caption{Background composition of the control region (CR) and signal region (SR).}
\label{fig:NOvA_QEH-BkgndCR}
\end{figure}

\begin{figure}[!t]
\centering
\includegraphics[width=0.50\linewidth]{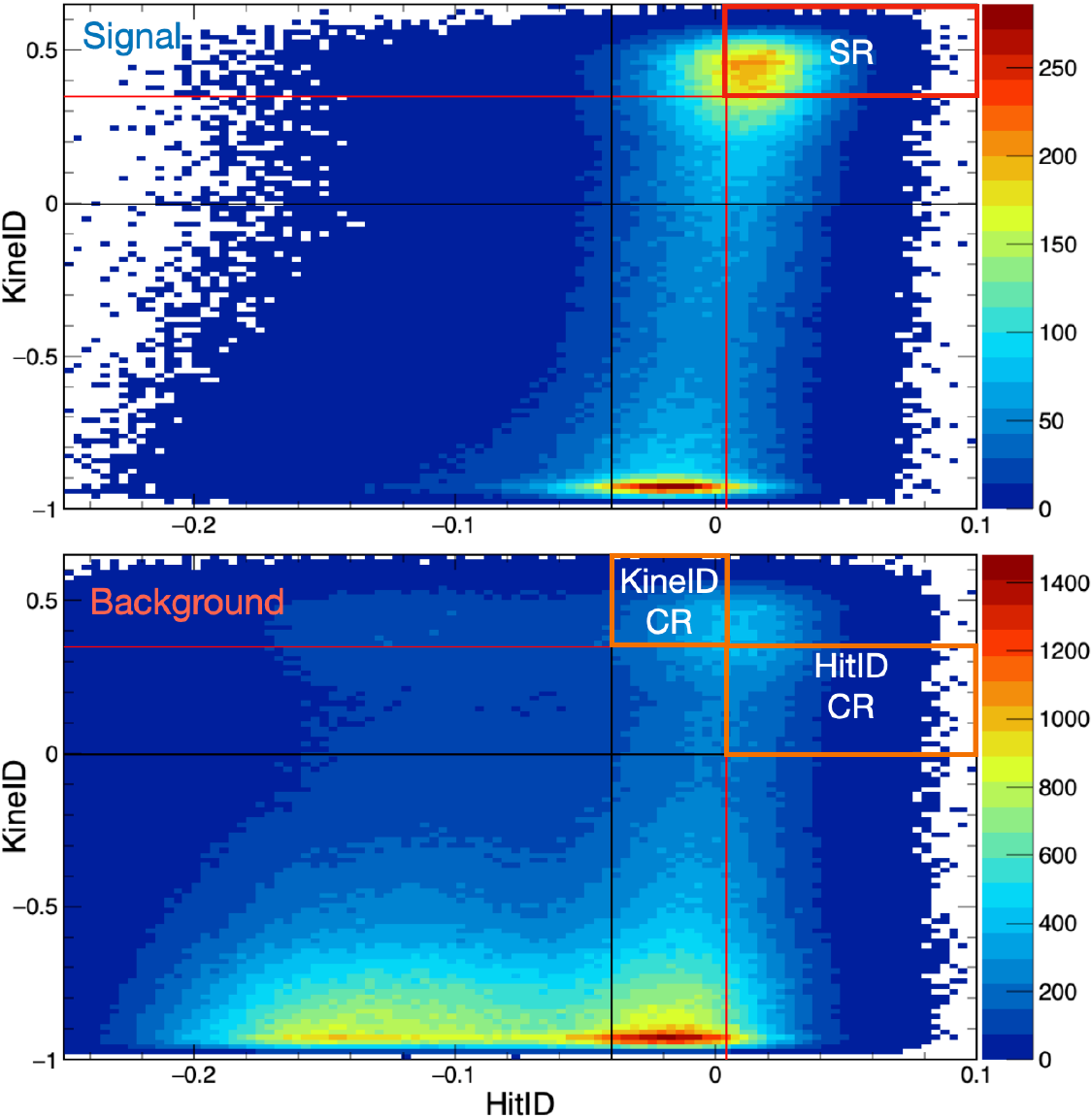}
\caption{Definition of the signal region and control regions in the two-dimensional plane of the topological (HitID) and kinematic (KineID) 
multivariate discriminants for uncontained events. The distributions are normalized to $1.2 \times 10^{21}$ POT.}
\label{fig:NOvA_QEH-2DplaneNNunc}
\end{figure}

The definition of the signal and control regions for contained events is presented in the main body of the paper (Fig.~\ref{fig:QEH-2DplaneNN}). 
The corresponding background compositions are shown in Fig.~\ref{fig:NOvA_QEH-BkgndCR}.  
Figure~\ref{fig:NOvA_QEH-2DplaneNNunc} shows the definition of the signal region and of the control regions in the plane of the HitID and KineID 
discriminants for uncontained events. The signal events grouped at negative values of KineID are characterized by an incorrect kinematics, due 
either to an underestimation of the muon energy or to a displaced detection point of the neutron candidate. The latter effect is also present for 
a small fraction of contained events (Fig.~\ref{fig:QEH-2DplaneNN}). Figure~\ref{fig:NOvA_QEH-EfficiencyPurity} shows the selection efficiencies 
and purities as a function of the antineutrino energy. The average values are given in the main body of the paper. 

\begin{figure}[!t]
\centering
\includegraphics[width=0.50\linewidth]{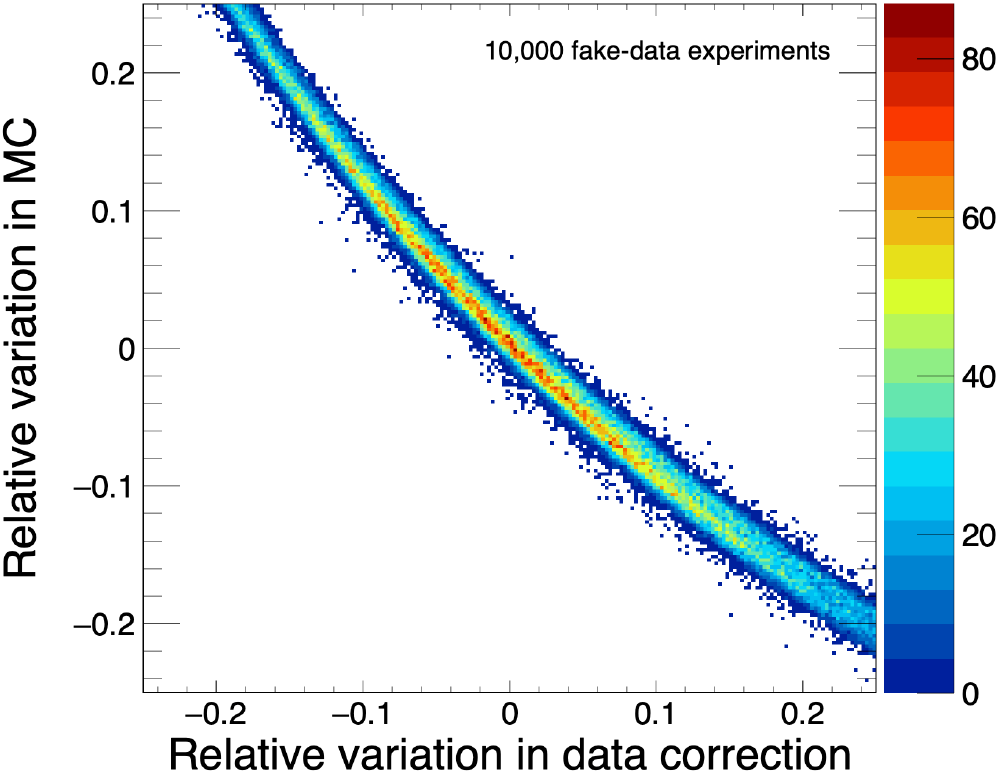}
\caption{Correlation between the relative variation in the MC backgrounds and the corresponding data corrections in Eq.~\eqref{eq:DScorr} from 10,000 fake-data experiments with random variations in the neutrino interaction models, in the neutron propagation model, and in the detector model. }
\label{fig:NOvA_QEH-EfficiencyPurity}
\label{fig:NOvA_QEH-MCvsKfactor}
\end{figure}

The main body of the paper describes the procedure determining the data corrections, as well as their use to constrain systematic variations 
in the MC (Fig.~\ref{fig:NOvA_QEH-DataSimulator}). Figure~\ref{fig:NOvA_QEH-MCvsKfactor} shows the correlation between the relative variation 
in the MC backgrounds and the corresponding data corrections from 10,000 fake-data experiments with random variations in the neutrino interaction models, 
in the neutron propagation model, and in the detector model. The data corrections are anticorrelated with the MC variations, as according to 
Eq.~\eqref{eq:DScorr} their product is ideally constant, up to the statistical accuracy of the control regions.

\begin{figure}[!t]
\centering
\includegraphics[width=1.00\linewidth]{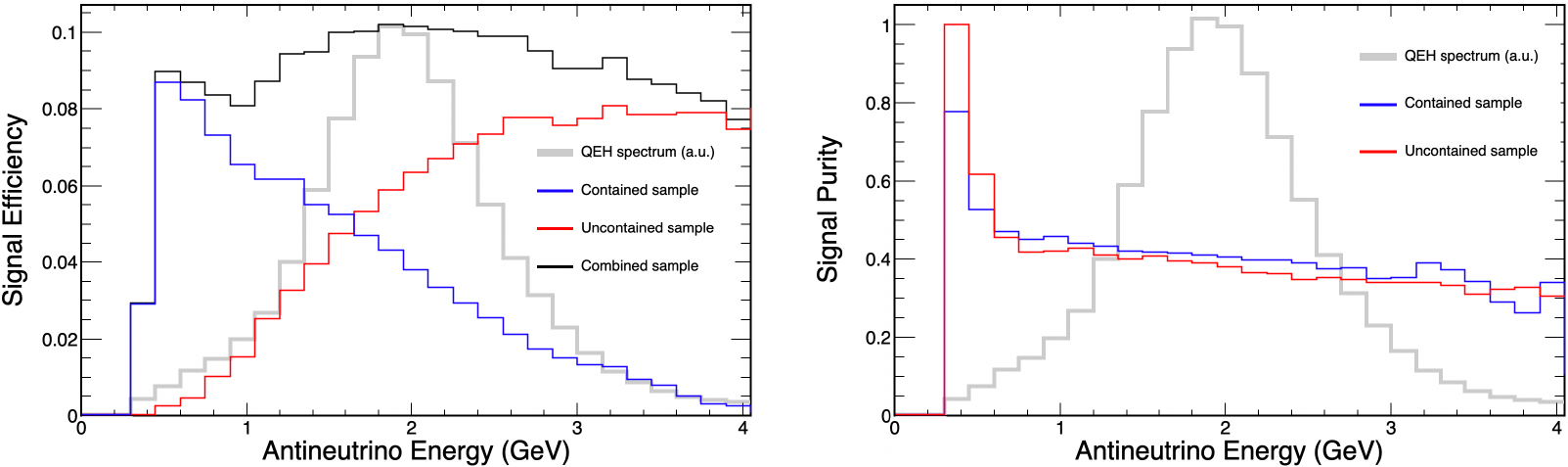}
\caption{Efficiencies (left plot) and purities (right plot) as a function of the antineutrino energy.}
\label{fig:NOvA_QEH-EfficiencyPurity}
\end{figure}

\end{document}

%% file: novaHelastic2026.tex
\newcommand{\ANL}{Argonne National Laboratory, Argonne, Illinois 60439, 
USA}
\newcommand{\Bandirma}{Bandirma Onyedi Eyl\"ul University, Faculty of 
Engineering and Natural Sciences, Engineering Sciences Department, 
10200, Bandirma, Balıkesir, Turkey}
\newcommand{\ICS}{Institute of Computer Science, The Czech 
Academy of Sciences, 
182 07 Prague, Czech Republic}
\newcommand{\IOP}{Institute of Physics, The Czech 
Academy of Sciences, 
182 21 Prague, Czech Republic}
\newcommand{\Atlantico}{Universidad del Atlantico,
Carrera 30 No.\ 8-49, Puerto Colombia, Atlantico, Colombia}
\newcommand{\BHU}{Department of Physics, Institute of Science, Banaras 
Hindu University, Varanasi, 221 005, India}
\newcommand{\UCLA}{Physics and Astronomy Department, UCLA, Box 951547, Los 
Angeles, California 90095-1547, USA}
\newcommand{\Caltech}{California Institute of 
Technology, Pasadena, California 91125, USA}
\newcommand{\CUSB}{Central University of South Bihar,
NH-120, Gaya Panchanpur Road, Post Fatehpur, 
Gaya, Bihar, 824 236, India}
\newcommand{\Cochin}{Department of Physics, Cochin University
of Science and Technology, Kochi 682 022, India}
\newcommand{\Charles}
{Charles University, Faculty of Mathematics and Physics,
 Institute of Particle and Nuclear Physics, Prague, Czech Republic}
\newcommand{\Cincinnati}{Department of Physics, University of Cincinnati, 
Cincinnati, Ohio 45221, USA}
\newcommand{\CSU}{Department of Physics, Colorado 
State University, Fort Collins, CO 80523-1875, USA}
\newcommand{\CTU}{Czech Technical University in Prague,
Brehova 7, 115 19 Prague 1, Czech Republic}
\newcommand{\Dallas}{Physics Department, University of Texas at Dallas,
800 W. Campbell Rd. Richardson, Texas 75083-0688, USA}
\newcommand{\DallasU}{University of Dallas, 1845 E 
Northgate Drive, Irving, Texas 75062 USA}
\newcommand{\Delhi}{Department of Physics and Astrophysics, University of 
Delhi, Delhi 110007, India}
\newcommand{\JINR}{Joint Institute for Nuclear Research,  
Dubna, Moscow region 141980, Russia}
\newcommand{\Erciyes}{
Department of Physics, Erciyes University, Kayseri 38030, Turkey}
\newcommand{\FNAL}{Fermi National Accelerator Laboratory, Batavia, 
Illinois 60510, USA}
\newcommand{\FSU}{Florida State University, Tallahassee, Florida 32306, USA}
\newcommand{\UFG}{Instituto de F\'{i}sica, Universidade Federal de 
Goi\'{a}s, Goi\^{a}nia, Goi\'{a}s, 74690-900, Brazil}
\newcommand{\Guwahati}{Department of Physics, IIT Guwahati, Guwahati, 781 
039, India}
\newcommand{\Harvard}{Department of Physics, Harvard University, 
Cambridge, Massachusetts 02138, USA}
\newcommand{\Homi}{Homi Bhabha National Institute, Training School Complex,
 Anushakti Nagar, Mumbai 400094, India}
\newcommand{\Houston}{Department of Physics, 
University of Houston, Houston, Texas 77204, USA}
\newcommand{\IHyderabad}{Department of Physics, IIT Hyderabad, Hyderabad, 
502 205, India}
\newcommand{\Hyderabad}{School of Physics, University of Hyderabad, 
Hyderabad, 500 046, India}
\newcommand{\IIT}{Illinois Institute of Technology,
Chicago IL 60616, USA}
\newcommand{\Imperial}{Imperial College London, Department of Physics,
 London, United Kingdom}
\newcommand{\Indiana}{Indiana University, Bloomington, Indiana 47405, 
USA}
\newcommand{\INR}{Institute for Nuclear Research of Russia, Academy of 
Sciences 7a, 60th October Anniversary prospect, Moscow 117312, Russia}
\newcommand{\UIowa}{Department of Physics and Astronomy, University of Iowa, 
Iowa City, Iowa 52242, USA}
\newcommand{\ISU}{Department of Physics and Astronomy, Iowa State 
University, Ames, Iowa 50011, USA}
\newcommand{\Irvine}{Department of Physics and Astronomy, 
University of California at Irvine, Irvine, California 92697, USA}
\newcommand{\Jammu}{Department of Physics and Electronics, University of 
Jammu, Jammu Tawi, 180 006, Jammu and Kashmir, India}
\newcommand{\Lebedev}{Nuclear Physics and Astrophysics Division, Lebedev 
Physical 
Institute, Leninsky Prospect 53, 119991 Moscow, Russia}
\newcommand{\Magdalena}{Universidad del Magdalena, Carrera 32 No 22-08 Santa Marta, Colombia}
\newcommand{\MSU}{Department of Physics and Astronomy, Michigan State 
University, East Lansing, Michigan 48824, USA}
\newcommand{\Crookston}{Math, Science and Technology Department, University 
of Minnesota Crookston, Crookston, Minnesota 56716, USA}
\newcommand{\Duluth}{Department of Physics and Astronomy, 
University of Minnesota Duluth, Duluth, Minnesota 55812, USA}
\newcommand{\Minnesota}{School of Physics and Astronomy, University of 
Minnesota Twin Cities, Minneapolis, Minnesota 55455, USA}
\newcommand{\Mississippi}{University of Mississippi, University, Mississippi 38677, USA}
\newcommand{\NISER}{
National Institute of Science Education and Research, 
Bhubaneswar, Khurda, Odisha 752050, India}
\newcommand{\OSU}{Department of Physics, Ohio State University, Columbus,
Ohio 43210, USA}
\newcommand{\Oxford}{Subdepartment of Particle Physics, 
University of Oxford, Oxford OX1 3RH, United Kingdom}
\newcommand{\Panjab}{Department of Physics, Panjab University, 
Chandigarh, 160 014, India}
\newcommand{\Pitt}{Department of Physics, 
University of Pittsburgh, Pittsburgh, Pennsylvania 15260, USA}
\newcommand{\QMU}{Particle Physics Research Centre, 
Department of Physics and Astronomy,
Queen Mary University of London,
London E1 4NS, United Kingdom}
\newcommand{\RAL}{Rutherford Appleton Laboratory, Science 
and 
Technology Facilities Council, Didcot, OX11 0QX, United Kingdom}
\newcommand{\SAlabama}{Department of Physics, University of 
South Alabama, Mobile, Alabama 36688, USA} 
\newcommand{\Carolina}{Department of Physics and Astronomy, University of 
South Carolina, Columbia, South Carolina 29208, USA}
\newcommand{\SDakota}{South Dakota School of Mines and Technology, Rapid 
City, South Dakota 57701, USA}
\newcommand{\SMU}{Department of Physics, Southern Methodist University, 
Dallas, Texas 75275, USA}
\newcommand{\Stanford}{Department of Physics, Stanford University, 
Stanford, California 94305, USA}
\newcommand{\Sussex}{Department of Physics and Astronomy, University of 
Sussex, Falmer, Brighton BN1 9QH, United Kingdom}
\newcommand{\Syracuse}{Department of Physics, Syracuse University,
Syracuse NY 13210, USA}
\newcommand{\Tennessee}{Department of Physics and Astronomy, 
University of Tennessee, Knoxville, Tennessee 37996, USA}
\newcommand{\Texas}{Department of Physics, University of Texas at Austin, 
Austin, Texas 78712, USA}
\newcommand{\Tufts}{Department of Physics and Astronomy, Tufts University, Medford, 
Massachusetts 02155, USA}
\newcommand{\UCL}{Physics and Astronomy Department, University College 
London, 
Gower Street, London WC1E 6BT, United Kingdom}
\newcommand{\Virginia}{Department of Physics, University of Virginia, 
Charlottesville, Virginia 22904, USA}
\newcommand{\WSU}{Department of Mathematics, Statistics, and Physics,
 Wichita State University, 
Wichita, Kansas 67260, USA}
\newcommand{\WandM}{Department of Physics, William \& Mary, 
Williamsburg, Virginia 23187, USA}
\newcommand{\Wisconsin}{Department of Physics, University of 
Wisconsin-Madison, Madison, Wisconsin 53706, USA}
\newcommand{\deceased}{Deceased.}
\affiliation{\ANL}
\affiliation{\Atlantico}
\affiliation{\Bandirma}
\affiliation{\BHU}
\affiliation{\Caltech}
\affiliation{\Charles}
\affiliation{\CUSB}
\affiliation{\Cincinnati}
\affiliation{\Cochin}
\affiliation{\CSU}
\affiliation{\CTU}
\affiliation{\Delhi}
\affiliation{\Erciyes}
\affiliation{\FNAL}
\affiliation{\FSU}
\affiliation{\UFG}
\affiliation{\Guwahati}
\affiliation{\Homi}
\affiliation{\Houston}
\affiliation{\Hyderabad}
\affiliation{\IHyderabad}
\affiliation{\IIT}
\affiliation{\Imperial}
\affiliation{\Indiana}
\affiliation{\ICS}
\affiliation{\INR}
\affiliation{\IOP}
\affiliation{\UIowa}
\affiliation{\ISU}
\affiliation{\Irvine}
\affiliation{\JINR}
\affiliation{\Magdalena}
\affiliation{\MSU}
\affiliation{\Duluth}
\affiliation{\Minnesota}
\affiliation{\Mississippi}
\affiliation{\NISER}
\affiliation{\OSU}
\affiliation{\Panjab}
\affiliation{\Pitt}
\affiliation{\QMU}
\affiliation{\SAlabama}
\affiliation{\Carolina}
\affiliation{\SMU}
\affiliation{\Sussex}
\affiliation{\Syracuse}
\affiliation{\Texas}
\affiliation{\Tufts}
\affiliation{\UCL}
\affiliation{\Virginia}
\affiliation{\WSU}
\affiliation{\WandM}
\affiliation{\Wisconsin}

\author{S.~Abubakar}
\affiliation{\Erciyes}

\author{M.~A.~Acero}
\affiliation{\Atlantico}

\author{B.~Acharya}
\affiliation{\Mississippi}

\author{P.~Adamson}
\affiliation{\FNAL}









\author{N.~Anfimov}
\affiliation{\JINR}


\author{A.~Antoshkin}
\affiliation{\JINR}


\author{E.~Arrieta-Diaz}
\affiliation{\Magdalena}

\author{L.~Asquith}
\affiliation{\Sussex}


\author{A.~Aurisano}
\affiliation{\Cincinnati}






\author{N.~Balashov}
\affiliation{\JINR}

% ...

\collaboration{The NOvA Collaboration}
\noaffiliation